\documentclass{emulateapj}

\slugcomment{ApJ -- Received 2005 March 23; Accepted 2005 September 17}

\shorttitle{Lyman Break Galaxy Clustering}
\shortauthors{Kashikawa et al.}

\begin{document}


\title{Clustering of Lyman Break Galaxies at $z=4$ and $5$ in The Subaru Deep Field: Luminosity Dependence of The Correlation Function Slope\altaffilmark{1}}


\author{
Nobunari Kashikawa\altaffilmark{2,3}, 
Makiko Yoshida\altaffilmark{4}, 
Kazuhiro Shimasaku\altaffilmark{4,5}, 
Masahiro Nagashima\altaffilmark{6}, 
Hideki Yahagi\altaffilmark{4}, 
Masami Ouchi\altaffilmark{7}, 
Yuichi Matsuda\altaffilmark{8}, 
Matthew A. Malkan\altaffilmark{9}, 
Mamoru Doi\altaffilmark{10}, 
Masanori Iye\altaffilmark{2,3}, 
Masaru Ajiki\altaffilmark{11}, 
Masayuki Akiyama\altaffilmark{12}, 
Hiroyasu Ando\altaffilmark{2}, 
Kentaro Aoki\altaffilmark{12}, 
Hisanori Furusawa\altaffilmark{12}, 
Tomoki Hayashino\altaffilmark{13}, 
Fumihide Iwamuro\altaffilmark{14}, 
Hiroshi Karoji\altaffilmark{14}, 
Naoto Kobayashi\altaffilmark{10}, 
Keiichi Kodaira\altaffilmark{15}, 
Tadayuki Kodama\altaffilmark{2}, 
Yutaka Komiyama\altaffilmark{2}, 
Satoshi Miyazaki\altaffilmark{12}, 
Yoshihiko Mizumoto\altaffilmark{2}, 
Tomoki Morokuma\altaffilmark{10}, 
Kentaro Motohara\altaffilmark{10}, 
Takashi Murayama\altaffilmark{11}, 
Tohru Nagao\altaffilmark{2,16}, 
Kyoji Nariai\altaffilmark{17}, 
Kouji Ohta\altaffilmark{8}, 
Sadanori Okamura\altaffilmark{4,5}, 
Toshiyuki Sasaki\altaffilmark{12}, 
Yasunori Sato\altaffilmark{2}, 
Kazuhiro Sekiguchi\altaffilmark{12}, 
Yasuhiro Shioya\altaffilmark{11}, 
Hajime Tamura\altaffilmark{13}, 
Yoshiaki Taniguchi\altaffilmark{11}, 
Masayuki Umemura\altaffilmark{18}, 
Toru Yamada\altaffilmark{2}, and 
Naoki Yasuda\altaffilmark{19}
}




\altaffiltext{1}{Based on data collected at the Subaru Telescope, which is operated by the National Astronomical Observatory of Japan.}
\altaffiltext{2}{Optical and Infrared Astronomy Division, National Astronomical Observatory, Mitaka, Tokyo 181-8588, Japan; kashik@zone.mtk.nao.ac.jp}
\altaffiltext{3}{Department of Astronomy, School of Science, Graduate University for Advanced Studies, Mitaka, Tokyo 181-8588, Japan.}
\altaffiltext{4}{Department of Astronomy, University of Tokyo, Hongo, Tokyo 113-0033, Japan.}
\altaffiltext{5}{Research Center for the Early Universe, University of Tokyo, Hongo, Tokyo 113-0033, Japan.}
\altaffiltext{6}{Department of Physics, Graduate School of Science, Kyoto University, Sakyo-ku, Kyoto 606-8502, Japan.}
\altaffiltext{7}{Space Telescope Science Institute, 3700, San Martin Drive, Baltimore, MD 21218.}
\altaffiltext{8}{Department of Astronomy, Graduate School of Science, Kyoto University, Kyoto 606-8502, Japan.}
\altaffiltext{9}{Department of Physics and Astronomy, University of California, Los Angeles, CA 90095.}
\altaffiltext{10}{Institute of Astronomy, University of Tokyo, Mitaka, Tokyo 181-8588, Japan.}
\altaffiltext{11}{Astronomical Institute, Graduate School of Science, Tohoku University, Aramaki, Aoba, Sendai 980-8578, Japan.}
\altaffiltext{12}{Subaru Telescope, National Astronomical Observatory of Japan, 650 North A'ohoku Place, Hilo, HI 96720.}
\altaffiltext{13}{Research Center for Neutrino Science, Graduate School of Science, Tohoku University, Aramaki, Aoba, Sendai 980-8578, Japan.}
\altaffiltext{14}{Department of Astronomy, Graduate School of Science, Kyoto University, Sakyo-ku, Kyoto 606-8502, Japan.}
\altaffiltext{15}{Graduate University for Advanced Studies (SOKENDAI), Shonan Village, Hayama, Kanagawa 240-0193, Japan.}
\altaffiltext{16}{INAF --- Osservatorio Astrofisico di Arcetri, Largo Enrico Fermi 5, 50125 Firenze, Italy.}
\altaffiltext{17}{Department of Physics, Meisei University, 2-1-1 Hodokubo, Hino, Tokyo 191-8506, Japan.}
\altaffiltext{18}{Center for Computational Physics, University of Tsukuba, 1-1-1 Tennodai, Tsukuba 305-8571, Japan.}
\altaffiltext{19}{Institute for Cosmic Ray Research, University of Tokyo, Kashiwa, Chiba 277-8582, Japan.}

%


\begin{abstract}
We explored the clustering properties of Lyman break galaxies (LBGs) at $z=4$ and $5$ with an angular two-point correlation function on the basis of the very deep and wide Subaru Deep Field data.
We confirmed the previous result that the clustering strength of LBGs depends on the UV luminosity in the sense that brighter LBGs are more strongly clustered.
In addition, we found an apparent dependence of the correlation function slope on UV luminosity for LBGs at both $z=4$ and $5$.
More luminous LBGs have a steeper correlation function.
The bias parameter was found to be a scale-dependent function for bright LBGs, whereas it appears to be almost scale-independent for faint LBGs.
Luminous LBGs have a higher bias at smaller angular scales, which decreases as the scale increases.
To compare these observational results, we constructed numerical mock LBG catalogs based on a semianalytic model of hierarchical clustering combined with high-resolution $N$-body simulation, carefully mimicking the observational selection effects.
The luminosity functions and the overall correlation functions for LBGs at $z=4$ and $5$ predicted by this mock catalog were found to be almost consistent with the observation.
The observed dependence of the clustering on UV luminosity was not reproduced by the model, unless subsamples of distinct halo mass were considered.
That is, LBGs belonging to more massive dark halos had steeper and larger amplitude correlation functions. 
With this model, we found that LBG multiplicity in massive dark halos amplifies the clustering strength at small scales, which steepens the correlation function.
The hierarchical clustering model could therefore be reconciled with the observed luminosity dependence of the correlation function if there is a tight correlation between UV luminosity and halo mass.
Our finding that the slope of the correlation function depends on luminosity could be an indication that massive dark halos hosted multiple bright LBGs.

\end{abstract}






\section{Introduction}

The clustering strength of galaxies is one of the most fundamental measures in observational cosmology.
The large-scale structure observed today was formed by the gravitational growth of initial small density fluctuations.
The measurement of galaxy clustering and its evolution enables us to see the history of galaxy assembly, which is an essential process in the present-day standard paradigm of hierarchical galaxy formation.
However, it is not so straightforward that clustering measurements alone can provide a direct determination of essential cosmological parameters or the initial power spectrum.
This is because clustering measurements can be made only with observable galaxies, not with the invisible dark matter halos, whose mass dominates the universe and drives gravitational interaction.
The formation process of observable galaxies in these dark matter halos remains uncertain, and is an important outstanding problem in the interpretation of observed galaxy clustering.
The formation of galaxies depends strongly on complicated physical processes connected to the evolution of the baryonic components, such as gas cooling, star formation, supernova feedback in each dark halo, and galaxy mergers.
We have to infer how galaxies trace the underlying dark matter throughout their evolution with cosmic time, for which purpose both detailed observations and comparison with theoretical modeling are essential.

The gravitational clustering evolution of dark matter from specified initial conditions has been thoroughly investigated in the hierarchical formation picture using both $N$-body simulations and analytic methods.
Semi-analytic approaches have been successful in modeling complex galaxy formation processes.
Advances in computer technology have enabled precise cosmological predictions to be made based on the use of more robust frameworks with larger volumes and finer resolution.
Indeed, in the local universe, detailed comparisons of galaxy clustering for various subsets of galaxies have been made for these models and the precise observations provided by the Sloan Digital Sky Survey (SDSS; \citealp{zeh02}) and the Two Degree Field Galaxy Redshift Survey (2dFGRS; \citealp{nor02}).
In addition to the previously known dependence of correlation amplitude on luminosity, color, morphology, and spectral type, more detailed signatures such as the slight slope differences among different subsamples and a systematic departure from a single power law have been revealed \citep{zeh04, zeh05}.
A similarly precise measurement of clustering is required for the high-$z$ universe to make a detailed comparison with the model predictions, which is now feasible.

On the theoretical side, clustering measurements in the high-$z$ universe were initially thought to be a valuable tool in distinguishing cosmological models.
The observed strong clustering with high bias at $z>3$, however, can be easily reproduced in most of the cold dark matter (CDM) cosmologies with either semianalytic approaches \citep{bla04,wec01,som01,gov98,bau98} or hydrodynamic simulations \citep{wei02, kat99}.
Since the precise measurement by the {\it Wilkinson Microwave Anisotropy Probe} ({\it WMAP}) of the cosmic microwave background (CMB) has established an appropriate cosmological model, clustering measurements have instead turned out to be an effective tool for constraining the nature of LBGs themselves \citep{kat99}.
As in the case of the local universe, a detailed study of clustering properties and their dependence on high-$z$ galaxy properties provides hints for understanding what determines these properties and how galaxies trace the underlying dark matter distribution.

Multicolor selection technique has efficiently revealed an abundant population of high-$z$ star-forming galaxies, called Lyman break galaxies (LBGs; \citealp{ste96}), at $z>2$.
It has been shown that this high-$z$ population has a large clustering length ($r_0=3$---$6h^{-1}$Mpc; \citealp{gia98, fou03}) comparable to that of present-day bright galaxies.
This result suggested a comparatively large value of the bias factor at $z\sim3$, which has generally been interpreted as indirect evidence that LBGs reside in massive dark matter halos with $M>10^{11}M_\odot$.
\citet{gia01} found an additional important tendency that more luminous LBGs have higher clustering strengths.
This implies that the luminosity (in this case, UV luminosity, which corresponds to star formation rate) of LBGs is governed by the halo mass of the galaxy.
The same attempts to quantify the amount of clustering have been applied to LBGs at higher $z$ \citep{ouc04b}, and the strength was found to be almost constant over the range $z=3$---$5$, which means high-bias galaxy formation during early epochs.
The observed clustering strength constrains the dark halo mass of LBGs by comparison with either analytic treatments \citep{mou02} or semianalytic models \citep{bla04}.

However, the fields of view (FOVs) of these surveys to probe high-$z$ galaxies are generally small, except in a few cases \citep{ouc04a,fou03}.
The small FOV leads to a systematic underestimation of the clustering scale and limits a statistically robust analysis against cosmic variance, which crucially affects the estimate of the normalization of the correlation function.
Small-scale measurements also fail to provide a true estimate of the shape of the correlation function, which is generally assumed to be a single power law $\xi(r)=(r/r_0)^{-\gamma}$ with a slope parameter $\gamma=1.8$.
The slope parameter is much more loosely constrained for high-$z$ LBGs, $\gamma=2.0\pm0.7$ \citep{gia98, gia01}, compared with the local value of $\gamma=1.71\pm0.06$ \citep{nor01} or $\gamma=1.75\pm0.03$ \citep{zeh02}.
In addition, there is an important implication that some dark halos may harbor more than one LBG, as suggested by numerical simulations \citep{som01, wec01}.
If this is the case, then clustering analyses made using a small FOV are biased to probe dominantly for galaxy pairs in the same halo and will tend to miss the halo-halo pair contribution, which has a distinct clustering strength at large scales.

We carried out a clustering analysis exploiting the very deep and wide imaging data of the Subaru Deep Field (SDF), which has an effective area of $876$ arcmin$^2$.
Our previous study on LBG clustering in the SDF was presented in \citet{ouc04b}. 
The present paper reports an extensive analysis based on deeper images.
The deepness of the SDF allows us to explore the faint LBG population with $M_{UV}=-19+5$log$h_{70}$ at $z=4$ and $M_{UV}=-20+5$log$h_{70}$ at $z=5$ \citep{yos06}, which are comparable to the faintest LBGs at $z=3$ \citep{gia01}.
The SDF surveyed area is $5$ times larger than a single Great Observatories Origins Deep Survey (GOODS) field \citep{gia04} and $250$ times larger than the Hubble Deep Field (HDF).
The clustering analysis of the Canada-France deep fields survey \citep{fou03} was based on a fairly large contiguous field ($28\arcmin\times28\arcmin$); however, their limiting magnitude with a $4$ m telescope is $\sim2$ mag shallower than ours.
The details of the luminosity dependence of the galaxy clustering have been difficult to establish owing to the narrow luminosity range probed thus far in the high-$z$ universe.
In this study, we make a detailed analysis of the luminosity dependence of LBG clustering based on deep and wide SDF data.


Furthermore, we directly compare the observed angular correlation functions at $z=4$ and $5$ with theoretical predictions given by a semianalytic model $+$ $N$-body simulation based on the hierarchical formation scheme.
Mock LBG catalogs were compiled from the simulation data with the same selection criteria as in the real observational sample.
This is the first study to make such a comparison of LBG clustering properties at $z=4$ and $5$ between observations and the simulations, and our deep and wide data allow a precise comparison.

The outline of the paper is as follows.
In \S~2, we describe our LBG sample.
In \S~3, we briefly review the model that we used and show that it predicts the same abundances of LBGs as the observations at $z=4$ and $5$.
The sky distribution of our observed LBG sample is shown in \S~4.
We present our results on the clustering analysis of the sample in \S~5.
A comparison between the observational results and the model predictions is made in \S~6, where we also discuss the possible interpretation of our results.
Our results are discussed further in \S~7, and conclusions are presented in \S~8.

Throughout the paper, we analyze in the flat $\Lambda$CDM model: $\Omega_m=0.3$, $\Omega_\Lambda=0.7$, $\sigma_8=0.9$, and the spectrum parameter $\Gamma=\Omega_m h=0.21$, where the Hubble constant is defined as $H_0=70h_{70}$kms$^{-1}$Mpc$^{-1}$. 
These parameters are consistent with recent CMB constraints \citep{spe03}.
Magnitudes are given in the AB system.

\section{The Data and the LBG sample}

We used a set of very wide and deep multi-color imaging catalogs of the SDF\footnotemark[20].
\footnotetext[20]{The SDF images and catalogs are available at http://soaps.naoj.org/sdf/.}
The observations, the data processing, and the source detections are described in \citet{kas04}.
The $3\sigma$ limiting magnitudes in $2\arcsec$ apertures are $28.45$, $27.74$, $27.80$, $27.43$, and $26.62$ in $B$, $V$, $R$, $i'$, and $z'$, respectively.
The final co-added image has an effective area of $876$ arcmin$^2$ with a unified seeing size of $0^{\prime \prime}.9$ for each band.
The wide FOV enables us to probe large comoving volumes of $8.844\times10^6$ and $5.390\times10^6h_{70}^{-3}$Mpc$^3$ for our LBG samples at $\langle z \rangle\simeq4$ and $\simeq5$, respectively.
The absolute error of the astrometry was found to be as small as $0^{\prime \prime}.21$-$0^{\prime \prime}.27$.

We used the two high-$z$ LBG samples presented in \citet{yos06}, in which full details of sample selection, completeness, and contamination rates are described.
Here we briefly describe our LBG samples.
These samples were constructed through a color selection technique that distinguishes high-$z$ galaxies having the strong Lyman break on their spectral energy distributions (SEDs).
We used the ($B-R$) versus ($R-i'$) two-color plane and the ($V-i'$) versus ($i'-z'$) plane to identify LBGs at $\langle z \rangle\simeq4$ (z4LBGs) and LBGs at $\langle z \rangle\simeq5$ (z5LBGs), respectively.
The exact selection criteria of the z4LBG sample are
\begin{eqnarray}
B-R  & \geq & 1.2 \nonumber\\
R-i' & \leq & 0.7 \nonumber\\
B-R  & \geq & 1.6(R-i')+1.9,
\end{eqnarray}
and those of the z5LBG sample are
\begin{eqnarray}
V-i'  & \geq & 1.2 \nonumber\\
i'-z' & \leq & 0.7 \nonumber\\
V-i'  & \geq & 1.8(i'-z')+1.7 \nonumber\\
B     & >    & 3\sigma.
\end{eqnarray}
These criteria are basically the same as those of \citet{ouc04a} but extend to fainter magnitudes.
All magnitudes are corrected for Galactic extinction \citep{sch98}, and all colors are measured in a $2\arcsec$ aperture.
In cases where the magnitude of an object is fainter than the $1\sigma$ limiting magnitude, it is replaced with the $1\sigma$ limiting magnitude.
For the z4LBG sample, $4543$ objects down to $i'_{tot}=27.4$ met our criteria, and for the z5LBG sample $831$ objects down to $z'_{tot}=26.6$ met our criteria.  
We adopted the {\tt MAG\_AUTO} value obtained by SExtractor \citep{ber96} for the total magnitude.

The completeness and contamination of these LBG samples were evaluated by \citet{yos06} in a way similar to that of \citet{ouc04a}.
We carried out a Monte Carlo simulation creating artificial objects with typical high-$z$ galaxy SEDs calculated with the model of \citet{ka97}, as well as the intergalactic medium attenuation model of \citet{mad95}.
We assumed the same $E(B-V)$ distribution as that of the LBG samples of \citet{ouc04a}.
These artificial objects were distributed on the real SDF images, then detected and measured in the same manner as for the real observed data.  
The completeness as a function of magnitude and redshift was defined as the ratio of the number of objects that were detected and satisfied our LBG color criteria to the input number of artificial objects.
On the other hand, the contamination rate was evaluated with the photometric redshift catalog of the HDF-N by \citet{fur00}.
We again carried out a Monte Carlo simulation, creating artificial objects with the same redshift, magnitude, and color distributions as the HDF-N catalog.
The contamination rate was defined as the ratio of the number of nearby objects satisfying the LBG color criteria to the number of LBGs.

So far, $22$ objects from the z4LBG sample and $16$ objects from the z5LBG sample have been spectroscopically identified.
Based on the follow-up spectroscopy, we have found that all of the objects that meet our criteria were indeed identified as high-$z$ LBGs; i.e., the nominal contamination rate is $0\%$, although the spectroscopic sample is still small.
\citet{yos06} constructed another LBG sample at $z\sim5$ using the ($R-i'$) versus ($i'-z'$) selection; however, we do not analyze this sample, as it would have almost the same redshift distribution as the z5LBG sample selected in the ($V-i'$) versus ($i'-z'$) plane, but with a larger contamination rate.

\section{$\nu$GC: The Mock Galaxy Catalog}

\subsection{A General Outline of the $\nu$GC}
We used a numerical mock galaxy sample to compare the predicted clustering strength with that of observations.
The model that we used is called the $\nu$GC (numerical galaxy catalog; \citealp{nag05}), which is implemented as a combination of two approaches, one a high-resolution $N$-body simulation to follow the hierarchical merging history of dark halos, and the other a semianalytic method (SAM) to model some physical processes related to the evolution of the baryonic components in each dark halo.

The $N$-body simulation has a box size of $100h_{70}^{-1}$Mpc on a side and treats $512^3$ dark matter particles.
Selecting only halos containing at least $10$ dark matter particles, the $\nu$GC follows the dark halo masses as small as $3\times10^9$M$_\odot$, which is appropriate for this study, as it includes the very faint and small-mass high-$z$ galaxy population.
The merger trees are directly constructed by the $N$-body simulation assuming a power spectrum of initial density fluctuations in a Gaussian random field.
Although individual small-mass dark halos are resolved, the $\nu$GC nevertheless covers a large volume, having a virtual FOV of $0.66\times0.66$ deg$^2$, which is about twice the size of the SDF.
Therefore, we can compare with a high confidence the clustering strength of the observations and predictions.
The $\nu$GC has the highest mass resolution among currently available cosmological simulations.
For example, the $\nu$GC has a spatial resolution $1.7$ times higher than that of the ^^ ^^ Millennium Simulation" \citep{spr05}, although the box size is much smaller.

Complicated galaxy formation processes can be described by a number of simple analytic model equations.
In the SAM used for the $\nu$GC, luminosities and colors of model galaxies were calculated with a population synthesis technique using simple stellar populations computed by \citet{ka97} with the Salpeter initial mass function from $0.1$ to $60M_\odot$, and an internal dust extinction is evaluated from the slab models.
In addition to galaxy mergers driven by dynamical friction, collisional starbursts were included during major mergers so as to assemble a new galaxy by converting all of the available cold gas into stars.
The SAM of the $\nu$GC also takes into account radiative cooling, supernova feedback, dynamical responses to gas removal, and tidal stripping, which have been shown to increase the agreement between models and observations.

Full details of the model itself are presented in \citet{nag05}, where it is shown that the model reproduces well the local luminosity function, color distribution, \ion{H}{1} gas mass fraction, galaxy size distribution and so on.
Further methodological detail is described in a couple of relevant references, \citet{yah04} for the $N$-body simulation and \citet{nag04} for the semianalytic method.

The mock galaxy catalog that we used for this study was generated from the $\nu$GC to meet selection criteria almost identical to those of the SDF catalog.
The magnitudes were determined with the same seeing size and isophotal threshold as for each band of the observations.
The surface brightness distribution of $\nu$GC mock galaxies was confirmed to be consistent with the SDF observations.
The predictions for galaxy number counts in the optical bands were found to be completely consistent with those observed in the SDF \citep{kas04}.
The predicted redshift distribution of $K$-band selected galaxies was also found to be consistent with the SDF observations \citep{kas03}.

\subsection{The SDF LBG Mock Catalogs}

\begin{figure}
\epsscale{1.25}
\plotone{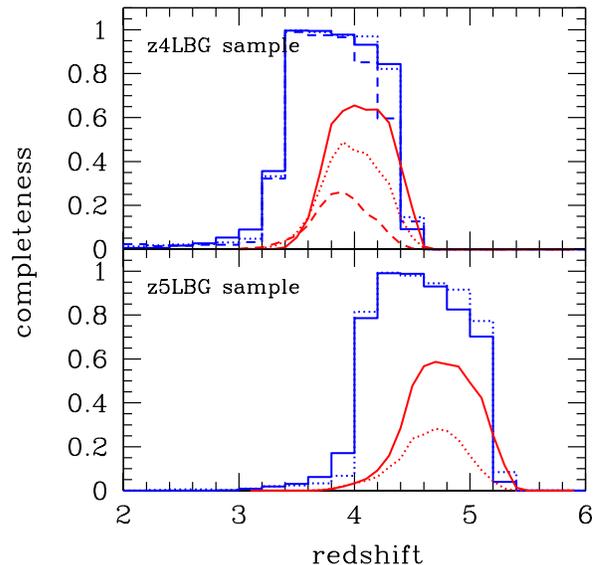}
\caption{Sample completeness of the $\nu$GC mock LBG sample. 
The blue histograms show the completeness of the color-selected (COL-selected) $\nu$GC LBG sample, while the red lines show the completeness estimate of the observed SDF LBG sample.
{\it Top:} z4LBGs in $i'\leq25.5$ ({\it solid line}), $25.5<i'\leq26.5$ ({\it dotted line}), and  $25.5<i'\leq27.4$ ({\it dashed line}).
{\it Bottom:} z5LBGs in $z'\leq25.75$ ({\it solid line}) and $25.75<z'\leq26.6$ ({\it dotted line}).
The completeness of the SDF LBG sample by \citet{yos06} was rebinned in magnitudes from the original plot taking the average with number weighting.
[{\it See the electronic edition of the Journal for a color version of this figure.}] 
\label{fig_compl}}
\epsscale{1.0}
\end{figure}

Based on the $\nu$GC described above, high-$z$ galaxy mock samples were extracted to mimic our observational z4LBG and z5LBG samples.
However, applying the same color criteria to the $\nu$GC as for the observed LBG samples does not guarantee that the mock samples have the same redshift and magnitude distributions as the observed samples. 
We did a couple of different estimates of the $\nu$GC predictions to see this uncertainty.

First, we applied the same color selection criteria to the $\nu$GC as for the actual observed SDF data to select LBGs at $z=4$ and $5$.
The sample that we constructed in this manner is defined as the ^^ ^^ COL-selected" sample.
In fact, we found that our selection criteria can be employed to isolate $z=4$ and $5$ galaxies correctly in the $\nu$GC catalog, as is the case for the observations.
This demonstrates that the semianalytic model imprinted on the $\nu$GC predicts the luminosity and color properties of these high-$z$ galaxies reasonably well.
Applying the color criteria, $27,456$ mock galaxies were extracted to be compared with the z4LBG sample, and $4712$ galaxies with the z5LBG sample.
Second, we randomly extracted numerical galaxies from the $\nu$GC so as to have completely the same selection function in redshift and magnitude as that observed.
The completeness estimate made by \citet{yos06} was regarded as the sample selection function.
The sample that we constructed in this manner is defined as the ^^ ^^ SF-selected" sample.
This contains $6653$ and $1690$ mock galaxies as the z4LBG and z5LBG samples, respectively.
The SF-selected sample has exactly the same redshift and magnitude distributions as observed, irrespective of their color criteria, while the COL-selected sample has exactly the same color criteria as observed, irrespective of their redshift and magnitude distribution.
These two differently selected mock galaxy samples have the same limiting magnitudes as those of the observations in the object detection bands, that is, $i'=27.4$ in the z4LBG sample and $z'=26.6$ in the z5LBG sample.

In Figure~\ref{fig_compl}, we show the completeness estimates of the COL-selected sample as a function of redshift.
These are almost the same as those of the observed sample \citep{yos06} apart from the normalization.
As seen in the estimate by \citet{yos06}, the actual observed catalog was affected by photometric errors, which severely decreased the sample completeness at fainter magnitudes.
This is not the case for the numerical simulation, where there is no photometric error.
The redshift distributions of the $\nu$GC LBG samples have slightly lower redshift extensions as compared to the estimates of \citet{yos06}.
Although the reason for this discrepancy is not clear at this time, such a difference is not unlikely because the completeness estimate of the observations is based on a galaxy SED model that is not exactly the same as that applied in the $\nu$GC.
On the other hand, the SF-selected sample has a selection function exactly identical to that of the observations, in which almost all ($\sim93\%$) the members are also included in the COL-selected sample.

\begin{figure}
\epsscale{1.25}
\plotone{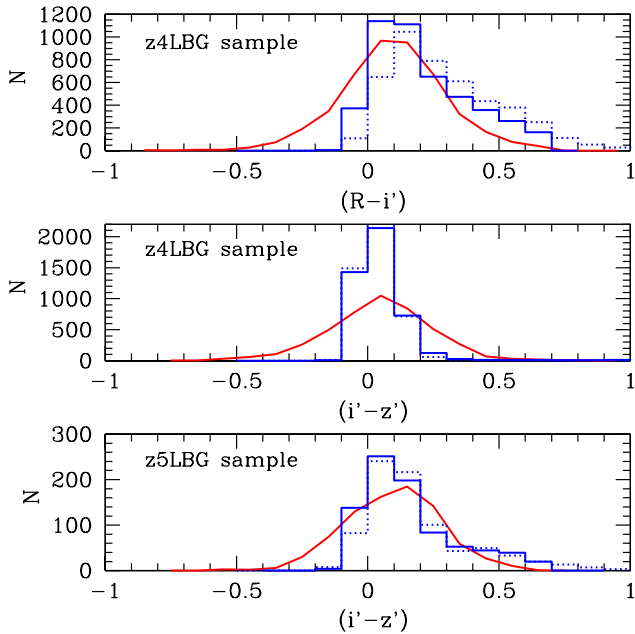}
\caption{Comparison of the rest-UV color distribution for LBGs at $z\sim4$ ({\it top and middle}) and $z\sim5$ ({\it bottom}) between SDF observations ({\it solid lines}) and the predictions of the $\nu$GC ({\it histograms}).
The solid histograms denote the color distribution of COL-selected $\nu$GC LBG samples, while the dotted histograms are those of SF-selected $\nu$GC LBG samples.
[{\it See the electronic edition of the Journal for a color version of this figure.}] 
\label{fig_col}}
\epsscale{1.0}
\end{figure}

Figure~\ref{fig_col} shows a comparison of the rest-UV color distribution between the observations and $\nu$GC predictions.
The colors, corresponding to the rest-UV continuum slope, are good measures of the internal dust attenuation of LBGs.
Note again that the observed colors have in general larger scatter than those of the simulation owing to photometric errors.
The observed and predicted color distributions show relatively good agreement with each other, except that the model prediction has a slightly elongated red tail in $(R-i')$ for z4LBGs and $(i'-z')$ for z5LBGs.
This agreement indicates that the dust extinction model implemented in the $\nu$GC works fairly well.
\citet{idz04} presented that the color selection applied to their semianalytic model reproduced well the observed completeness of the LBG sample at $z\sim4$ from the GOODS survey, although the $i_{775}-z_{850}$ color did not agree completely, which is probably due to a deficiency of the dust amount or an incorrect extinction adopted in the model.
Our SDF LBG sample was found to show almost the same $i'-z'$ color distribution as that of GOODS, and we confirmed that the $\nu$GC mock z4LBG sample has a similar $i'-z'$ distribution with $73\%$ and $75\%$ likelihoods on a Kolmogorov-Smirnov test for the COL and SF-selected samples, respectively.
The $\nu$GC reproduces the $i'-z'$ color of z4LBGs much better than that of \citet{idz04}; however, the color predictions of the $\nu$GC should still be affected by imperfect dust reddening, which also has a significant effect on the completeness estimate for the COL-selected sample.

\begin{figure}
\epsscale{1.25}
\plotone{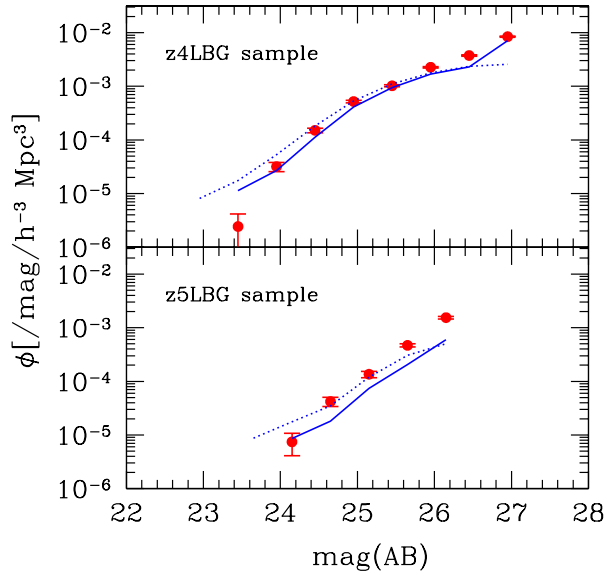}
\caption{Comparison of the luminosity functions for LBGs at $z\sim4$ ({\it top}) and $z\sim5$ ({\it bottom}) between SDF observations by \citet{yos06} ({\it circles with error bars}) and the predictions of the $\nu$GC ({\it lines}).
The solid lines denote the luminosity functions of COL-selected $\nu$GC LBG samples, while the dotted lines are those of SF-selected $\nu$GC LBG samples.
[{\it See the electronic edition of the Journal for a color version of this figure.}] 
\label{fig_lf}}
\epsscale{1.00}
\end{figure}


In Figure~\ref{fig_lf}, we show a comparison of the luminosity functions of LBGs at $z=4$ and $5$ between the observation and the model predictions of the $\nu$GC.
The observed luminosity functions were derived in \citet{yos06} in almost the same way as in \citet{ste99} to correct the selection function and incompleteness of the sample.
The effective comoving volume for each magnitude bin was calculated by convolving the comoving volume and the completeness as a function of the redshift shown in Figure~\ref{fig_compl}.
The contamination correction was applied to the samples as a function of magnitude.
In the predicted COL-selected samples, the luminosity function can be obtained in the same way as in the observations, although the effective volume was derived in the $\nu$GC on its own.
In contrast, for the SF-selected sample, in which the observational selection function was used, the luminosity function was determined simply by dividing the number counts of the sample by the effective volume derived by
\citet{yos06} for each magnitude bin.

The resulting luminosity functions of the COL-selected and SF-selected samples are denoted in Figure~\ref{fig_lf} by solid and dotted lines, respectively.
They are almost consistent with each other and with the observations as well; especially striking agreement can be seen in the z4LBG sample.
The agreement of the COL-selected sample with the observations means that the simulated galaxies in the $\nu$GC have realistic colors at high $z$, while the agreement between the SF-selected sample and observation verifies that the $\nu$GC predicts the same number density and redshift distribution as is observed at high $z$.
The SF-selected sample has a slightly shallower luminosity function than that of the COL-selected sample; however, our present results for LBG clustering show almost no difference between the COL-selected and SF-selected sample, as shown below.


\section{Sky Distribution}

\begin{figure}
\epsscale{.82}
\epsscale{1.2}
\plotone{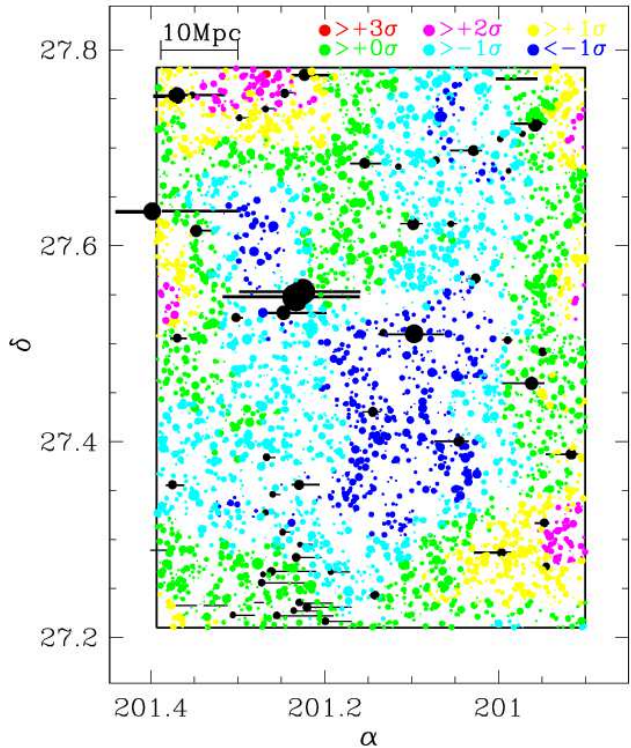}
\caption{Sky distribution of SDF z4LBG sample. 
Larger circles denote brighter objects in total $i'$-band magnitude.
The local overdensity, with the significance of each, is denoted by color as shown in the color legend.
The black shaded areas are masked regions in which detected objects were rejected due to low $S/N$.
North is up and east is to the left. 
[{\it See the electronic edition of the Journal for a color version of this figure.}] 
\label{fig_skyz4}}
\epsscale{1.0}
\end{figure}

\begin{figure}
\epsscale{.82}
\epsscale{1.2}
\plotone{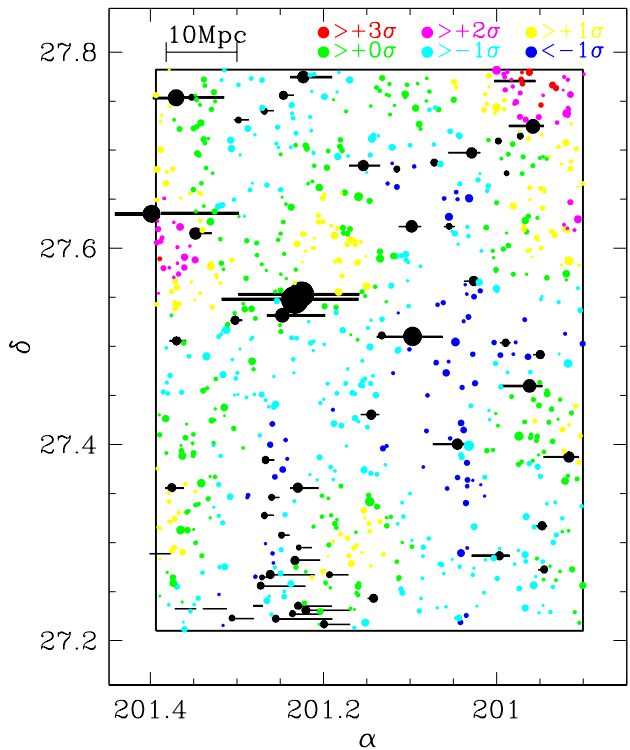}
\caption{Same sky distribution as in Figure~\ref{fig_skyz4}, but for the SDF z5LBG sample.
[{\it See the electronic edition of the Journal for a color version of this figure.}] 
\label{fig_skyz5}}
\epsscale{1.0}
\end{figure}

Figure~\ref{fig_skyz4} and Figure~\ref{fig_skyz5} show the sky distributions of our z4LBG and z5LBG samples, respectively.
Larger circles denote brighter objects in total $i'$-band and $z'$-band magnitude for the z4LBG and z5LBG samples, respectively.
The circle color represents the local surface number density of the sample around the object.
We measured the surface number density in a circle of $8h_{70}^{-1}$Mpc comoving radius around each sample object.
The mean value and the dispersion of the number density were derived by the statistics of $10,000$ randomly chosen positions over the image.
In Figure~\ref{fig_skyz4} and Figure~\ref{fig_skyz5}, the overdensities with $3$, $2$, $1$, $0$, $-1$, and $<-1\sigma$ significances are denoted with colors as indicated in the figures.
The black shaded areas are masked regions\footnotemark[21] where detected objects were rejected due to low $S/N$.
\footnotetext[21]{The masked regions are defined at the SDF data release web site, http://soaps.naoj.org/sdf/.}
It is strikingly apparent from these figures that galaxies are distributed inhomogeneously even in the $z=4$ and $5$ universe.
In the case of the z4LBG sample, the highest number density is $1.78$($h_{70}^{-2}$Mpc$^2$)$^{-1}$  which has a $2.9\sigma$ significant density excess, while the lowest density is $0.696$($h_{70}^{-2}$Mpc$^2$)$^{-1}$ with a $-1.1\sigma$ significant density drop.
In the z5LBG sample, the highest number density is $0.458$($h_{70}^{-2}$Mpc$^2$)$^{-1}$ ($3.5\sigma$ significance) and the lowest density is $0.092$($h_{70}^{-2}$Mpc$^2$)$^{-1}$ ($-2.2\sigma$).
Note that the number density is evaluated for the z4LBG and z5LBG samples with different limiting magnitudes down to $i'\leq27.4$ and $z'\leq26.6$, respectively.

\section{Clustering analysis of the SDF LBG sample}

\subsection{Angular Correlation Function}

We derived the angular two-point correlation functions (ACFs) $w(\theta)$ to estimate the clustering strength of our photometric galaxy sample.
We used the \citet{ls93} estimator of $w(\theta)$,
\begin{eqnarray}
w(\theta)=\frac{DD-2DR+RR}{RR},
\end{eqnarray}
where the $DD$, $DR$, and $RR$ denote the number of galaxy-galaxy, galaxy-random, and random-random pairs having angular separations between $\theta$ and $\theta+\delta\theta$.
We generated $100,000$ random points to reduce the Poisson noise in random pair counts and normalized $DD$, $DR$, and $RR$ to the total number of pairs in each pair count.
The random points were created with exactly the same boundary conditions as the SDF galaxy sample avoiding the mask regions where saturated stars dominate.
The ACF can be approximated by the power-law form given by
\begin{eqnarray}
w(\theta)=A_w \theta^{\beta},
\end{eqnarray}
and we quantified the ACF with these two parameters, amplitude $A_w$ and slope $\beta$.
The measured ACFs have to be corrected for the ^^ ^^ integral constraint" (IC) bias, which is caused by the uncertainty of the mean galaxy density estimated from the sample itself.
We estimated the IC as follows:
\begin{eqnarray}
{\rm IC}=\frac{1}{\Omega^2}\int\!\!\!\int w(\theta) d\Omega_1 d\Omega_2,
\end{eqnarray}
where $\Omega$ is the surveyed area.
Assuming that $w(\theta)$ is denoted by Eq.(4), IC is a function of $\beta$ and proportional to $A_w$.
For example, we derived IC$=0.00279A_w$ for $\beta=-0.8$.
The IC was derived for several $\beta$-values and was found to be a smooth function of $\beta$.
In this study, we used a different IC correction according to the best-fit value of $\beta$, although the maximum variance of IC is only $\delta {\rm IC}=0.0027$.
For the mock catalog, which has an FOV twice as wide as the SDF, we obtained a smaller IC variance $\delta {\rm IC}=0.0023$.


Contamination by foreground galaxies would change the amplitude of the ACF.
In this study, we assumed that these contaminating galaxies have a close to homogeneous distribution.
In this case, the true correlation amplitude $A_w$ is reduced by a factor of $(1-f_c)^2$, where $f_c$ is the contamination rate.
The $f_c$ of our LBG sample is estimated as a function of magnitude in \citet{yos06}, and amounts to at most $20\%$ .

On the other hand, the completeness of the sample should not affect the ACF estimate, provided that the completeness is diluted homogeneously over the FOV.
\citet{yos06} confirmed that the completeness distribution of our LBG sample has little variation over the SDF FOV and thus completeness can be assumed to be a function of magnitude only.
We therefore did not correct the derived ACF in any way to take account of the completeness.
Assuming the weak correlation limit, we estimated the Poissonian errors only \citep{ls93} for the ACF as
\begin{eqnarray}
\sigma_w(\theta)=\sqrt\frac{1+w(\theta)}{DD}.
\end{eqnarray}
Actually, this estimate was confirmed to follow the notation of Poissonian error in \citet{arn02} quite well and was found to be the dominant error contribution to the ACF up to $100^{\prime \prime}$.

The clustering strength parameter $r_0$, of the spatial correlation function $\xi=(r/r_0)^\gamma$, is a useful parameter for comparisons with other studies.
It can be derived from the ACF through Limber's equation \citep{pee80} if the redshift distribution of the sample is well known and $F(z)$, the redshift dependence of the clustering strength, is properly assumed.
In this study, we used the completeness estimate as shown in Figure~\ref{fig_compl} as the redshift distribution of the sample and made the same assumption for $F(z)$ as that in \citet{ouc04b}.

\begin{figure}
\epsscale{1.30}
\plotone{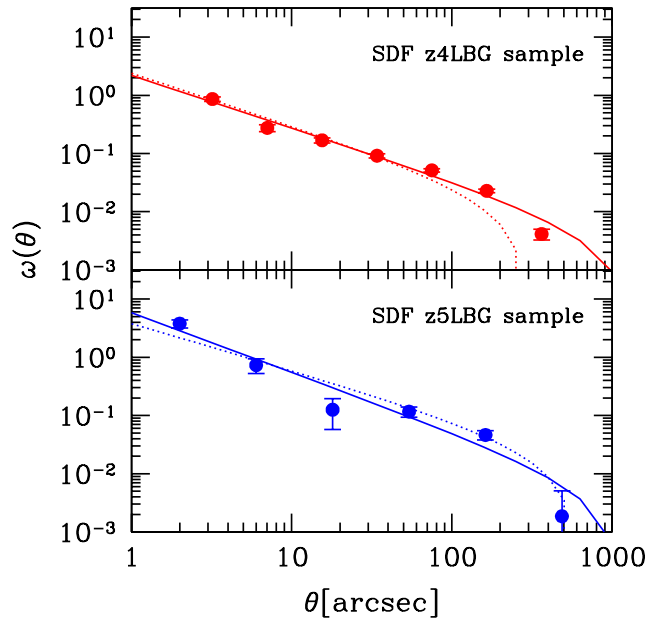}
\caption{ACFs of z4LBG ({\it top}) and z5LBG ({\it bottom}) samples.
The solid lines show the power-law fits with IC corrections. 
Error bars show the $1\sigma$ Poissonian errors.
The dotted lines show the ACF power law of the previous estimates in SDF by \citet{ouc04b}.
[{\it See the electronic edition of the Journal for a color version of this figure.}] 
\label{fig_acf_all}}
\epsscale{1.0}
\end{figure}

Figure~\ref{fig_acf_all} shows the ACF for our total z4LBG and z5LBG samples.
The power-law fitting derived from Eq.(4) is represented by the solid lines, and the derived parameters are listed in Table~\ref{tbl-acfparam}.
The ACF of the z5LBG sample shows a slightly steeper slope than that of the z4LBG sample.
For comparison, dotted lines show the previous ACF estimates in the SDF derived in \citet{ouc04b} with a fixed slope of $\beta=-0.8$. 
These are nearly consistent with the present results.

\subsection{Luminosity Dependence}

\begin{figure}
\epsscale{1.3}
\plotone{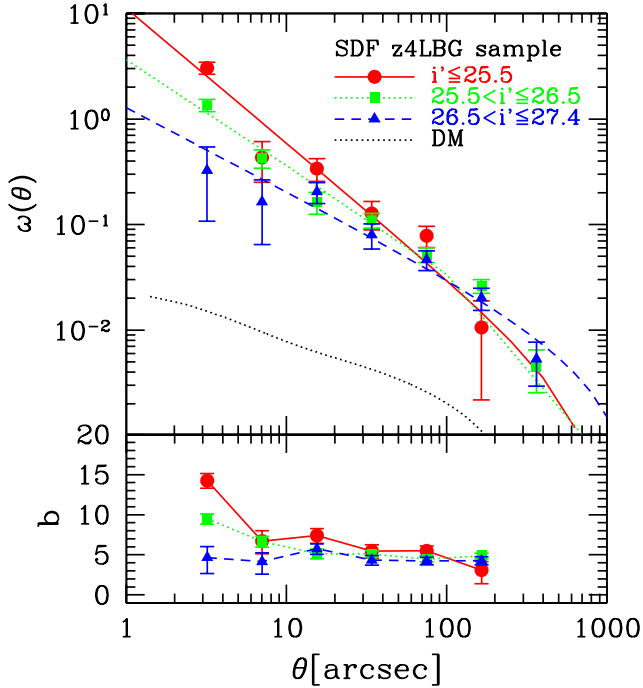}
\caption{Luminosity dependence of ACFs in the z4LBG sample.
The ACFs of $i'\leq25.5$, $25.5<i'\leq26.5$, and $26.5<i'\leq27.4$ are represented by symbols given in the legend.
The dotted line is the nonlinear ACF of dark matter at $z=4$ calculated with the same observational selection function.
{\it Bottom:} Our defined bias parameter $b(\theta)$ for each luminosity subsample with the same symbol and color as indicated in the top panel.
[{\it See the electronic edition of the Journal for a color version of this figure.}] 
\label{fig_macfz4}}
\epsscale{1.0}
\end{figure}

\begin{figure}
\epsscale{1.3}
\plotone{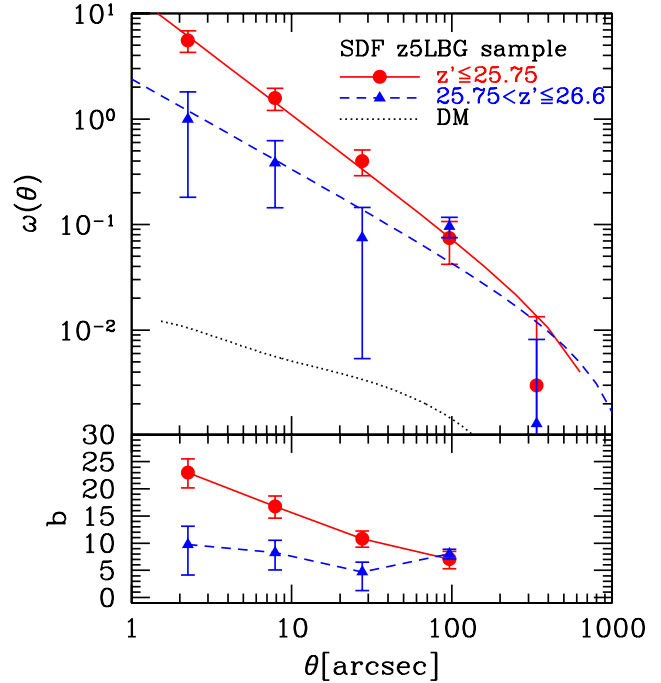}
\caption{Same as Figure~\ref{fig_macfz4}, but for the SDF z5LBG sample.
The ACFs of $z'\leq25.75$ and $25.75<z'\leq26.6$ are represented by symbols given in the legend.
[{\it See the electronic edition of the Journal for a color version of this figure.}] 
\label{fig_macfz5}}
\epsscale{1.0}
\end{figure}

Next, we investigated the UV luminosity dependence of the LBG ACF.
The sample was divided into three subsamples based on their $i'$-band magnitude and two based on $z'$-band magnitude for the z4LBG and z5LBG samples, respectively.
Note that these subsamples with different apparent magnitudes are taken to be subsamples of different rest UV luminosity because the sample galaxies have almost the same distance from us for each of the z4LBG and z5LBG samples.
The results are presented in Figure~\ref{fig_macfz4} and Figure~\ref{fig_macfz5}, and the best-fit parameters of the individual subsamples are given in Table~\ref{tbl-acfparam}.
We confirmed the previous results that the clustering strength of LBGs depends on the UV luminosity of the sample; that is, brighter LBGs are more strongly clustered.
The most remarkable result revealed in these figures is that not only the amplitude but also the slope of the LBG ACF has a strong dependence on luminosity in both the z4LBG and z5LBG samples, in the sense that brighter subsamples have higher ACF amplitudes and steeper ACF slopes.
In other words, more luminous LBGs have stronger clustering strengths at smaller scales.

\begin{figure}
\epsscale{1.2}
\plotone{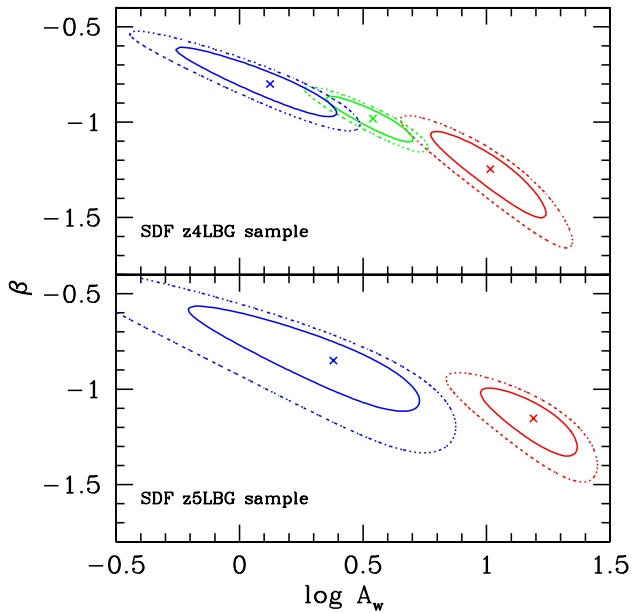}
\caption{Error ellipses of derived ACF parameters of $A_w$ and $\beta$ for luminosity subsamples of z4LBGs ({\it top}) and z5LBGs ({\it bottom}).
The line colors are identical to Figure~\ref{fig_macfz4} and Figure~\ref{fig_macfz5}.
Ellipses from left to right denote from fainter to brighter luminosity subsamples.
The solid and dotted ellipses are the $1\sigma$ and $3\sigma$ confidence levels.
[{\it See the electronic edition of the Journal for a color version of this figure.}] 
\label{fig_errcirc}}
\epsscale{1.0}
\end{figure}

To show the significance of the result, we plot the error contours for our two-parameter fits in Figure~\ref{fig_errcirc}.
The confidence levels for the fitting were computed based on Poissonian error statistics.
The subsamples with different UV luminosity have different ACF slopes at $1\sigma$ and greater significance, although this is weaker than that for the difference in ACF amplitudes.
Moreover, the robustness of our result, in which the clustering amplitude difference at small scales is essential, was evaluated by means of the blockwise bootstrap method \citep{por02}.
We first divided the whole SDF area into $3\times4 (=12)$ subregion samples of equal area.
We then constructed $100$ artificial bootstrap samples, each consisting of $12$ subsamples randomly chosen from the whole subsample set with allowance for duplications.
The ACF was calculated for each of these artificial bootstrap samples, and an error for each bin at small scales was estimated from the fluctuation among the $100$ bootstrap samples.
The amplitude differences at small scales in luminosity-distinct subsamples were reproduced by this method for both the z4LBG and z5LBG samples.
Bootstrap errors were found to be almost comparable to the Poissonian error, except for the smallest scale bin of the z5LBG sample, in which the bootstrap error was $3$ times higher than the Poissonian, due to small number statistics.

It has been suggested by previous studies of LBG clustering \citep{gia01, ouc04b, ade05} that the amplitude of the ACF has a strong dependence on UV luminosity; however, these studies almost always assumed a fixed slope of $\beta=-0.8$, mainly due to the statistical uncertainty in small LBG samples.
In this study based on a larger LBG sample with wider magnitude coverage, we find the first evidence that the ACF slope has a dependence on the UV luminosity of LBGs.
\citet{fou03} concluded that their clustering measurements for a $z\sim3$ LBG sample are broadly consistent with a power-law slope of $-0.8$ over all magnitudes; however, their brighter subsample ($I_{AB}=20.0-23.5$) has a steeper slope relative to that of the fainter subsample ($I_{AB}=23.5-24.5$), although the significance of this difference is low.
In \citet{gia01}, on the other hand, the deep $z\sim3$ LBG sample of the HDF was found to have a slightly steeper slope than that of bright ground-based samples, which is inconsistent with our result.
However, this trend is quite weak within the errors, as they mentioned, once both cosmic variance due to their small FOV and the different filter bands and different criteria for choosing the LBG sample for the HDF and the ground-based data are taken into account.


The bias parameter, which is defined as the ratio of the clustering amplitudes of galaxies to that of dark matter, is normally measured on large scales of $8$Mpc.
To see the bias parameter variance with scale, we define the bias parameter $b(\theta)$ as a function of scale as 
\begin{eqnarray}
b(\theta)=\sqrt\frac{w_g(\theta)}{w_{DM}(\theta)},
\end{eqnarray}
where $w_g(\theta)$ and $w_{DM}(\theta)$ are the ACFs of galaxies and dark matter, respectively.
We calculated $w_{DM}$ based on the nonlinear power spectrum of \citet{smi03} with the same observational selection function as shown in Figure~\ref{fig_compl}.
In Figure~\ref{fig_macfz4} and Figure~\ref{fig_macfz5} $w_{DM}$ is plotted in dotted lines, and $b(\theta)$ is plotted in the lower panels.
For the bright LBG sample, $b(\theta)$ has an apparently increasing trend with decreasing scale, whereas for the faint LBG sample, $b(\theta)$ has almost no dependence on scale.
The scale dependence of the bias parameter for the bright LBG sample is consistent with \citet{ham04}, which is based on our previous SDF result in which bright LBGs were dominant.
The scale-dependent bias at small scales is a generic prediction of hierarchical models independent of the epoch and of the model details \citep{col99}.
We compare this scale dependency with the model predictions in the next section.

\begin{figure}
\epsscale{.85}
\epsscale{1.1}
\plotone{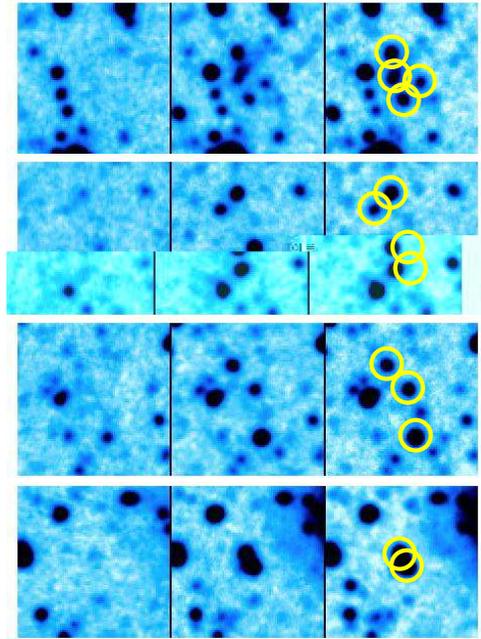}
\caption{Example of multiple LBGs at z=4 on SDF.
Objects indicated by circles in the right panels are the SDF z4LBG sample.
$B$-, $R$-, and $i'$-band images are shown from left to right.
Each image is $15\arcsec$ on a side.
North is up and east is to the left.
[{\it See the electronic edition of the Journal for a color version of this figure.}] 
\label{fig_mlbgz4}}
\epsscale{1.0}
\end{figure}

\begin{figure}
\epsscale{.85}
\epsscale{1.1}
\plotone{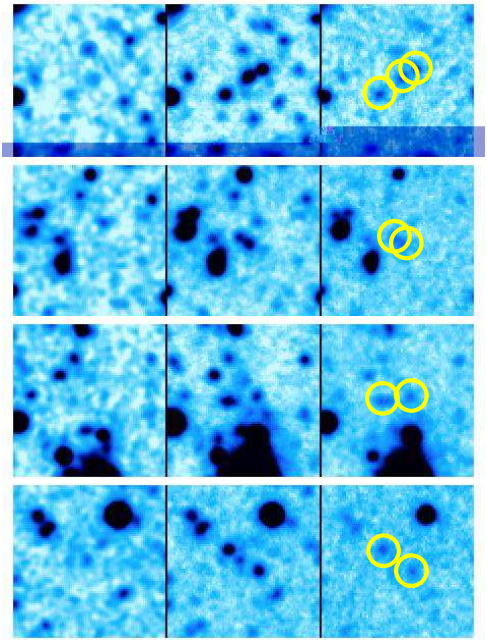}
\caption{Same as Figure~\ref{fig_mlbgz4}, but for z=5.
$V$-, $i'$-, and $z'$-band images are shown from left to right.
[{\it See the electronic edition of the Journal for a color version of this figure.}] 
\label{fig_mlbgz5}}
\epsscale{1.0}
\end{figure}

We found that the luminosity dependence of LBG clustering strength is more significant on smaller ($<10^{\prime \prime}$) scales.
The virial radius of a $10^{12}M_\odot$ halo at $z=4$-$5$ is roughly $\sim380h_{70}^{-1}$kpc in a comoving scale ($\sim75 h_{70}^{-1}$kpc in a physical scale), which corresponds to $10^{\prime \prime}$-$12^{\prime \prime}$.
Therefore, a large clustering difference was found on scales almost inside the halo virial radius.
This scale dependence of LBG clustering recalls an implication that massive dark halos should contain multiple bright LBGs.
In fact, we found pairs or multiple systems of bright LBGs with small ($\sim10^{\prime \prime}$) separations in our sample, as shown in Figure~\ref{fig_mlbgz4} and Figure~\ref{fig_mlbgz5}.
Note that we are measuring only projected separations so far; further spectroscopy is required to confirm that these are indeed spatially close systems.
It is quite often assumed that there is a one-to-one correspondence between LBGs and halos, that is, a single dark matter halo hosts a single galaxy, and some studies have suggested that this is correct \citep{ade98, gia01}.
However, our results suggest the possibility of a large contribution from galaxy pairs in the same halo at high luminosities and on smaller scales.
If two or more galaxies reside in a dark massive halo, these pairs naturally boost the ACF amplitude at smaller scales.
It would be interesting to investigate the halo occupation distribution (HOD) formalism that statistically describes the relation of galaxies within a dark matter halo; however, such an analysis is beyond the scope of this paper.
Instead, we discuss these interpretations further in the next section in the course of direct comparisons with semianalytic model predictions.


\subsection{Color Dependence}

We also investigated the color dependence of the ACF features of our LBG samples.
In this case, we use color to mean the steepness of the UV continuum slope, which is correlated with dust attenuation $E(B-V)$ \citep{meu99}.
SED fitting to synthesized stellar population models \citep{pap02, sha01} suggests that more luminous LBGs are dustier.
The redder LBGs are expected to be intrinsically brighter.
To see the difference in clustering strength between dusty and less dusty LBGs of the same luminosity, we made z4LBG subsamples with $(i'-z')>0$ and $\leq0$ at $i'\leq26.0$.
At $i'>26.0$, the shallower limiting magnitude of the $z'$ band as compared to the $i'$ band produces a color limit for the sample, which severely decreases the $(i'-z')\leq0$ subsample; thus, we did not include these faint galaxies.
We tried to make a color subsample for the z5LBG sample using the $zb$ and $zr$ data presented in \citet{shi05}, although the $zr$-band data have too shallow a limiting magnitude to make subsamples without the color cut bias.

The best-fit ACF parameters for the color subsample of z4LBG are presented in Table~\ref{tbl-acfparam}.
Although the dusty subsample with red colors of $(i'-z')>0$ has a slightly stronger clustering amplitude, it is within the $1\sigma$ errors of the same value for the $(i'-z')\leq0$ subsample.
Note that we here assume that redder LBGs are dustier; however, the UV continuum slope also depends on the stellar age, and no strong evidence of large dust content in LBGs has been found from submillimeter photometry \citep{web03}.
Taking account of a possible correlation between age and extinction \citep{sha01}, no clustering difference between different $E(B-V)$ subsamples could refute the hypothesis that the age difference causes the ACF slope difference. 
We discuss this later in \S~7.3.

\section{Comparison with The Model}

\subsection{The overall ACF}

In this section, we compare our observed estimates of LBG clustering at $z=4$ and $5$ with the predictions of the hierarchical clustering model based on the mock SDF/LBG catalogs.
As described in \S~2, we constructed two LBG mock catalogs, the COL-selected and SF-selected samples.
The selection function of the COL-selected sample has almost no dependence on magnitude, as shown in Figure~\ref{fig_compl}, which differs from the real observation.
Since the clustering strength depends strongly on the luminosity distribution of the sample, as seen in the previous section, it is essential to match the luminosity distribution between the model and observations.
We therefore randomly picked galaxies from the COL-selected sample so that the sample had the same normalization in selection function as observed for each magnitude bin.
The SF-selected sample, by definition, had already been drawn randomly from the original $\nu$GC catalog so as to have the same selection function in magnitude and in redshift as observed.
We iterated this random resampling procedure $30$ times, then derived correlation functions from the average value.
The error for each scale bin, $\sigma_w(\theta)$, was determined by the larger of either the Poissonian error estimated by Eq.(6) or the scatter in the random resampling.

\begin{figure}
\epsscale{1.3}
\plotone{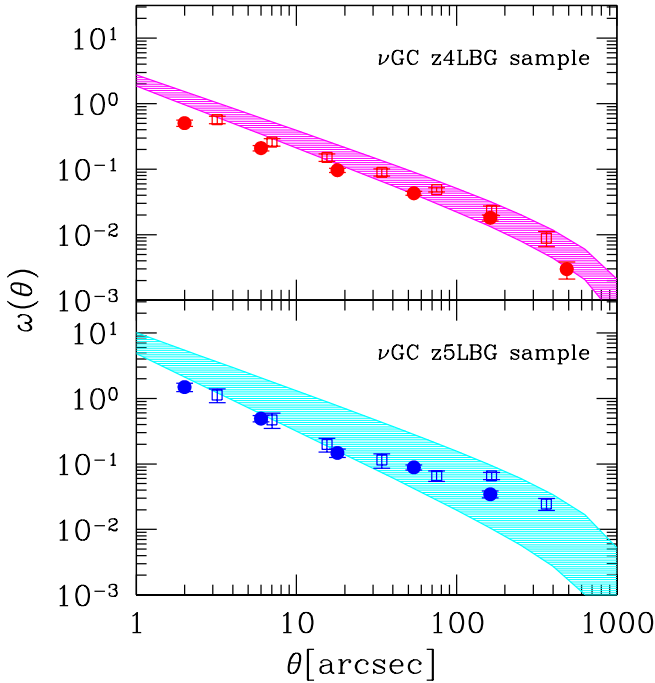}
\caption{Model-predicted ACFs by $\nu$GC mock samples for z4LBGs ({\it top}) and z5LBGs ({\it bottom}).
The filled circles and open squares denote the COL-selected and SF-selected samples, respectively.
The shaded regions show the observed ACF of the SDF total sample with $1\sigma$ tolerance.
[{\it See the electronic edition of the Journal for a color version of this figure.}] 
\label{fig_nugcall}}
\epsscale{1.0}
\end{figure}

Figure~\ref{fig_nugcall} shows the ACF for our total z4LBG and z5LBG samples, and the derived power-law fitting parameters are listed in Table~\ref{tbl-acfparamnugc}.
The filled circles and open squares denote the COL-selected and SF-selected samples, respectively, and the shaded regions show the observed ACF with $1\sigma$ tolerance.
Both of the mock samples show reasonable agreement with the observations at $z=4$ and $5$, although the observed ACFs have a somewhat steeper slope than the predictions with a slightly stronger strength in amplitude on small scales ($<10^{\prime \prime}$) for the z4LBG sample.
The two model predictions have almost the same clustering strength on all scales.
As a whole, we conclude that the $\nu$GC can reproduce the observed ACFs at $z=4$ and $5$.

\subsection{Luminosity and halo mass dependence of the ACF}

The most acute interest is in whether the model can reproduce the ACF slope dependence on the luminosity, as seen in our observed LBG samples.
We made luminosity-distinct mock subsamples in the same magnitude ranges as in the observed subsamples.
In this case, we did not apply a random culling to the COL-selected sample as was done in the previous section in order to match the selection function with that of the observed sample.
This is because the magnitude difference of the selection function in each luminosity subsample is negligible.
We used the model galaxy magnitudes, taking into account internal dust extinction, for comparison with the
observations.

\begin{figure}
\epsscale{.95}
\epsscale{1.3}
\plotone{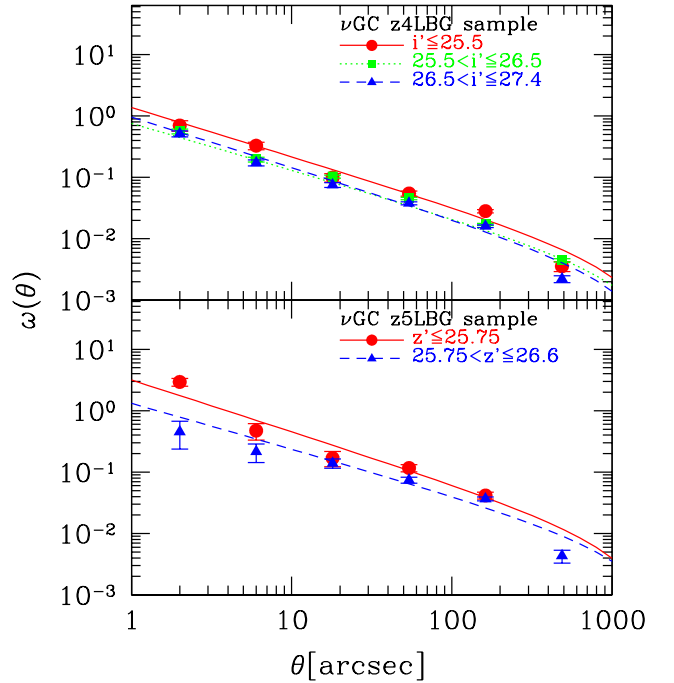}
\caption{Luminosity dependence of ACFs in the $\nu$GC prediction of the COL-selected sample for z4LBGs ({\it top}) and z5LBGs ({\it bottom}).
The symbols are identical to those in Figure~\ref{fig_macfz4} and Figure~\ref{fig_macfz5}.
[{\it See the electronic edition of the Journal for a color version of this figure.}] 
\label{fig_nugclum}}
\epsscale{1.0}
\end{figure}

Figure~\ref{fig_nugclum} shows the ACF luminosity dependence of the $\nu$GC prediction for the COL-selected sample.
In contrast to our expectations, there are almost no clustering differences among different luminosity subsamples.
Slight differences of amplitude can be seen, although they are much smaller than for the observations.
We confirmed that this null result is virtually the same in the case of the SF-selected sample.
\citet{wec01} also concluded that semianalytic models generally show a weak dependence of LBG clustering on luminosity.

\begin{figure}
\epsscale{.95}
\epsscale{1.3}
\plotone{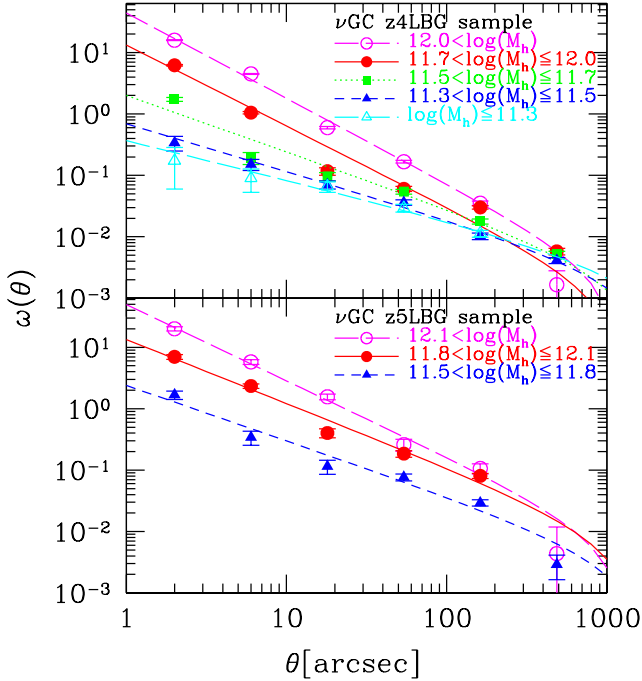}
\caption{Halo mass dependence of ACFs in $\nu$GC for z4LBGs ({\it top}) and z5LBGs ({\it bottom}).
The halo mass subsamples of $12.0<$log$(M_h/M_\odot)$, $11.7<$log$(M_h/M_\odot)\leq12.0$, $11.5<$log$(M_h/M_\odot)\leq11.7$, $11.3<$log$(M_h/M_\odot)\leq11.5$, and log$(M_h/M_\odot)\leq11.3$ are represented by symbols as shown in the legend, from top to bottom for z4LBGs.
The mass subsamples of $12.1<$log$(M_h/M_\odot)$, $11.8<$log$(M_h/M_\odot)\leq12.1$, and $11.5<$log$(M_h/M_\odot)\leq11.8$ are represented by symbols as shown in the legend, from top to bottom for z5LBGs.
[{\it See the electronic edition of the Journal for a color version of this figure.}] 
\label{fig_nugcmh}}
\epsscale{1.0}
\end{figure}

A natural expectation might be that the observed clustering difference is caused by dynamical mass differences of the sample galaxies.
To investigate this more straightforwardly, we made subsamples of the LBG mock catalog according to the dark halo mass, $M_h$, determined from the $N$-body simulation.
In this case, we accounted only for the limiting magnitude of the sample and did not correct the completeness for each halo mass subsample.
In the further analysis below, we used only the COL-selected sample with a few contributions of low-$z$ contamination removed to reduce uncertainties.
Figure~\ref{fig_nugcmh} shows the ACFs for subsamples with different halo mass ranges.
The clustering differences are clearly noticeable.
The ACF for higher halo masses has a steeper slope and stronger amplitudes at smaller scales, which is in excellent agreement with our observational results.

\begin{figure}
\epsscale{.82}
\epsscale{1.20}
\plotone{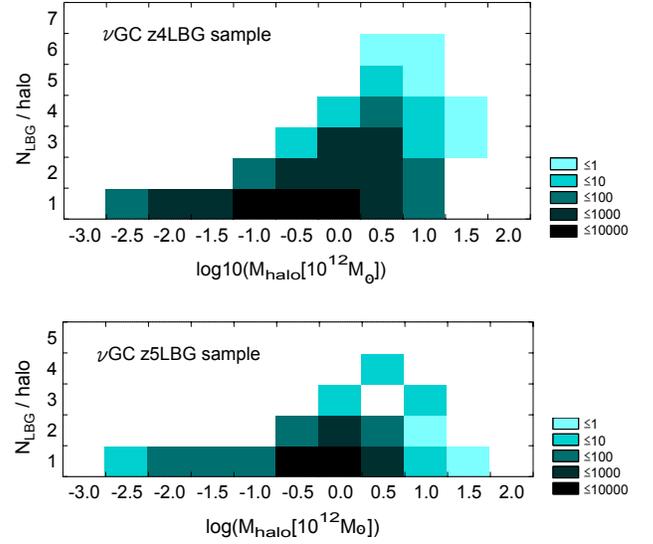}
\caption{
Two-dimensional histograms of the dark halo mass and the LBG occupation number in a halo predicted by the $\nu$GC for z4LBGs ({\it top}) and z5LBGs ({\it bottom}).
Each color level shows the number of galaxies as shown in the color legend; the darker shaded region shows greater numbers of samples.
The halo mass $M_{halo}$ is shown in units of $10^{12}M_\odot$.
[{\it See the electronic edition of the Journal for a color version of this figure.}] 
\label{fig_nlbg}}
\epsscale{1.0}
\end{figure}

Although it is in accordance with the expectation that more massive galaxies show stronger clustering strengths, it is rather surprising that the $\nu$GC model reproduces the ACF slope difference according to halo mass.
Most hierarchical clustering models, which adopt a variety of star formation history scenarios, predict that a more massive dark halo contains a larger number of galaxies with $N_{\rm LBG}\propto M_h^{0.7}-M_h^{0.8}$ \citep{ben00, wec01}.
This is also confirmed by smoothed particle hydrodynamics simulation \citep{ber03}.
This relation is also expected in the $\nu$GC, which takes account of collisional star formation.
Figure~\ref{fig_nlbg} illustrates the $\nu$GC predicted occupation numbers of LBGs ($N_{LBG}$) residing in a dark matter halo.
Note that in Figure~\ref{fig_nlbg}, we consider only halos harboring at least one galaxy, so there could be halos containing no galaxies that are not cataloged in the $\nu$GC.
Although most of the halos have only one LBG, some massive halos have multiple LBGs, and more massive halos tend to show higher multiplicity.
This halo mass dependence of LBG multiplicity explains our finding that more luminous LBGs have higher clustering amplitudes at scales as small as the virial radius of the dark halos.
These multiple galaxies in a dark massive halo can amplify the correlation strength effectively on small scales.
Furthermore, the ACF of LBGs in less massive halos could be dominated by the contribution of halo-halo correlation at all the scales, which is consistent with our result of a scale-independent bias parameter for faint LBGs, as can be seen in Figure~\ref{fig_macfz4} and Figure~\ref{fig_macfz5}.
Figure~\ref{fig_nugcmh} ({\it top}, z4LBG sample) also predicts that the halo-mass distinct ACFs have almost no differences for the low-mass halo subsample below $10^{11.5}M_{\odot}$, where one-to-one correspondence can be applied for all scales.
Our observed luminosity dependence of the ACF slope is qualitatively consistent with the picture that massive dark halos harbor multiple LBGs, which contradicts the assumption of one-to-one correspondence in the bright LBG description.

\subsection{ACFs of 1halo-1LBG sample}

\begin{figure}
\epsscale{1.2}
\plotone{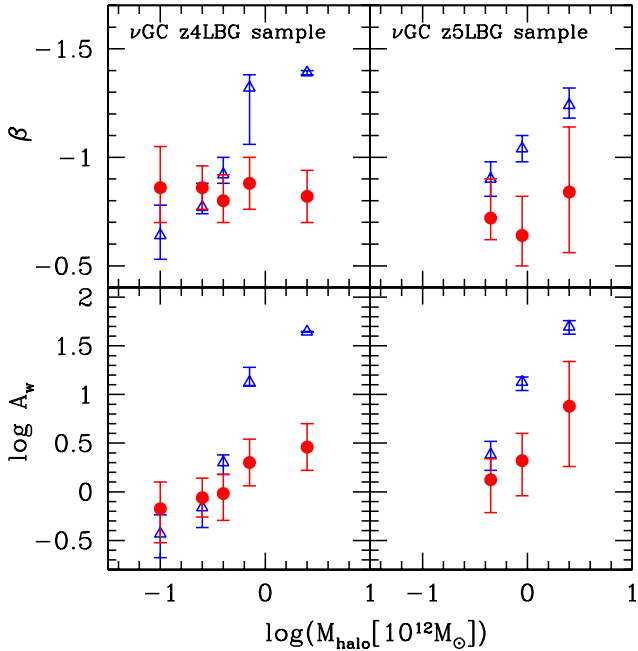}
\caption{
ACF parameters, $\beta$ and $A_w$, of the halo-mass distinct subsamples in the $\nu$GC 1halo-1LBG catalog.
The circles denote the ACF parameters of the 1halo-1LBG sample, while the triangles denote those of the original mock catalog derived in Figure~\ref{fig_nugcmh} for comparison.
{\it Left:} ACF slope $\beta$ ({\it top}) and ACF amplitude $A_w$ ({\it bottom}) of the z4LBG sample as a function of dark halo mass.
{\it Right:} Same as left, but for the z5LBG sample.
[{\it See the electronic edition of the Journal for a color version of this figure.}] 
\label{fig_1g1h}}
\epsscale{1.0}
\end{figure}

In the $\nu$GC mock catalog, host dark halos can be identified for each galaxy.
To further examine the hypothesis that the steep ACF slope of the massive dark halos as seen in Figure~\ref{fig_nugcmh} is caused by LBG multiplicity in individual halos, we made an artificial one-to-one correspondence mock catalog, in which all satellite galaxies were removed and the central brightest galaxy was left when a halo has multiple galaxies.
In this experimental ^^ ^^ 1halo-1LBG" mock sample, every halo has only one galaxy.
The ACF parameters were then derived for each mass-distinct subsample extracted from the 1halo-1LBG sample.
The results are listed in Table~\ref{tbl-acfparamnugc}, and the comparison with those derived from the original mock catalog is shown in Figure~\ref{fig_1g1h}.
In all cases, we found no excess and sometimes a deficit of ACF amplitudes at small scales ($<10^{\prime \prime}$) for the 1halo-1LBG sample.
In the case of deficit amplitudes, we derived the ACF parameters by fitting a power law at only the scales larger than $10^{\prime \prime}$.
The upper panels of Figure~\ref{fig_1g1h} apparently reveal that the ACF slope, $\beta$, does not differ among the halo-mass distinct subsamples in the 1halo-1LBG catalog.
The slopes of the ACFs of the 1halo-1LBG subsample coincide with each other around $\beta\sim-0.8$, which corresponds to that of the small halo mass subsample [log$(M_h/M_\odot)\la11.5$] in the original mock catalog.
Even a high-mass subsample now has the same ACF slope as that of a low-mass subsample in the 1halo-1LBG catalog.
This simple test clearly demonstrates that LBG multiplicity in massive dark halos causes the ACF amplitude excess at small scales, and halo occupation number plays an important role in the small-scale clustering of LBGs.

Figure~\ref{fig_macfz4} and Figure~\ref{fig_macfz5} show that brighter LBGs have a higher amplitude even at scales ($10^{\prime \prime}-100^{\prime \prime}$) larger than the typical virial radius of dark halos ($10^{\prime \prime}-12^{\prime \prime}$ for a $10^{12}M_\odot$ halo).
Can this luminosity dependence of clustering amplitude on larger scales be explained only by the halo-halo contribution?
The lower panels of Figure~\ref{fig_1g1h} show the ACF amplitude $A_w$ of halo-mass distinct subsamples in the 1halo-1LBG catalog.
They still show amplitude differences, although the significance of difference is smaller than that of the original catalog, which means that the semianalytic model $+$ $N$-body simulation predicts that more massive dark halos are intrinsically strongly clustered.
We can conclude that the model predicts that the slope of the halo-halo contribution to the ACF is always the same irrespective of halo mass, whereas the amplitude at larger scales is higher for more massive LBGs.
The observed trend in which more luminous galaxies have steeper and higher clustering strength can be interpreted as a synergy of two effects: first that more massive halos contain multiple LBGs, and second that more massive halos are strongly clustered.


\section{Discussion}

\subsection{UV Luminosity - Halo Mass Relation}

\begin{figure}
\epsscale{1.2}
\plotone{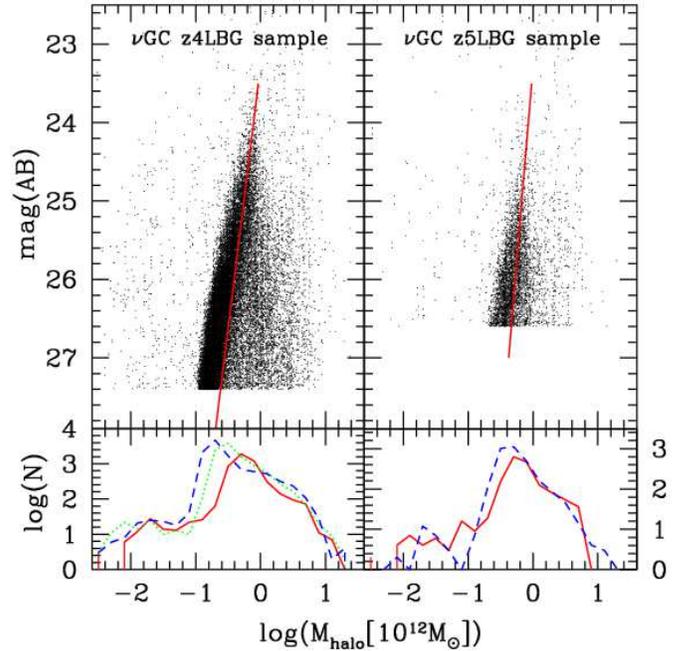}
\caption{Prediction of $\nu$GC regarding the UV luminosity dependence of galactic halo mass.
The magnitudes quoted are for the $i'$ band in the z4LBG sample ({\it top left}) and the $z'$ band in the z5LBG sample ({\it top right}).
The straight lines show the linear fit to the mean ridge of the distribution.
{\it Bottom:} Halo mass histograms for each luminosity subsample with the same lines as indicated in Figure~\ref{fig_macfz4} and Figure~\ref{fig_macfz5}.
[{\it See the electronic edition of the Journal for a color version of this figure.}] 
\label{fig_magmh}}
\epsscale{1.0}
\end{figure}

The predicted ACF behavior of the mass-selected sample is significantly in contrast to that of the luminosity-selected sample as seen in Figure~\ref{fig_nugclum}.
The hierarchical clustering model reproduces well the observed LBG clustering difference in the halo mass-distinct sample but not that in the luminosity-distinct sample.
To see this more closely, we show the relation between halo mass and galaxy magnitude predicted by the $\nu$GC in Figure~\ref{fig_magmh}.
The luminosity is nearly proportional to the halo mass, although the relation has rather large scatter.
The red line shows the linear fit to connect the mean ridges of the distribution.
The lower panels of Figure~\ref{fig_magmh} show the halo mass distribution for each of the $\nu$GC luminosity subsamples.
Although the distribution peak shifts slightly to lower mass as luminosity decreases, the overall features of the halo mass distribution are unchanged with luminosity.

Based on the mean ridge in Figure~\ref{fig_magmh} without considering the large scatter, the hierarchical clustering model predicts the mass of halos hosting z4LBGs of each UV luminosity as follows:
\begin{eqnarray}
10^{11.7}M_\odot<M_h<10^{12.0}M_\odot \  (23.5<i'\leq25.5),\nonumber\\
10^{11.5}M_\odot<M_h<10^{11.7}M_\odot \  (25.5<i'\leq26.5),\nonumber\\
10^{11.3}M_\odot<M_h<10^{11.5}M_\odot \  (26.5<i'\leq27.4).
\end{eqnarray}
A slightly weaker luminosity dependence of halo mass is seen for z5LBGs:
\begin{eqnarray}
10^{11.8}M_\odot<M_h<10^{12.1}M_\odot \  (24.0<z'\leq25.75),\nonumber\\
10^{11.5}M_\odot<M_h<10^{11.8}M_\odot \  (25.75<z'\leq26.6).
\end{eqnarray}
These values are comparable to the estimate for LBGs at $z=3$ by \citet{ade05} and smaller than the estimate by \citet{bla04}, and almost consistent with the estimate by \citet{ouc04b} for z4LBGs at $i'=24.5$, although the $\nu$GC predicts a shallower $M_h-L$ relation.

Taking into account these correspondences of UV luminosity and halo mass in comparing the ACFs between the observations in Figure~\ref{fig_macfz4} and Figure~\ref{fig_macfz5} and the model prediction in Figure~\ref{fig_nugcmh}, the observed clustering result for each luminosity subsample is reconciled roughly well with the predictions for each halo mass subsample.
A tighter relation between halo mass and UV luminosity would improve the consistency of the ACF dependence on luminosity between the observations and model predictions.
This conjecture is consistent with the idea that the observed slope difference of ACFs is due to differences in the occupation number of LBGs in a halo.
In Figure~\ref{fig_nugcmh}, multiple LBGs in the same halo were inevitably in the same mass-distinct subsample.
However, this is not always the case for luminosity-distinct subsamples in a simulated LBG sample having a UV luminosity and halo mass relation with a large scatter; i.e., multiple LBGs in the same halo could be separated into different luminosity-distinct subsamples.
On the other hand, observed luminosity-distinct subsamples have the same clustering tendency as the mass-distinct subsamples of the model.
Therefore, the true LBGs are supposed to have a tighter relation between halo mass and UV luminosity than the model prediction.
In turn, the halo mass plays a fundamental role in the star formation activity in LBGs.

The collisional starburst model that the $\nu$GC takes into account may predict a looser correlation between halo mass and UV luminosity that is sensitive to star formation history and dependent on the details of the merger process.
Moreover, the large scatter of the halo-mass---luminosity relation is a generic feature of the hierarchical clustering model suggested by \citet{wec01}.
\citet{som01} also pointed out a weak luminosity dependence on halo mass in the semianalytic model for $z\sim0$ galaxies.
This is a consequence of different star formation histories in different halos. 
\citet{som01} raise the possibility that the dust extinction increases the mass-to-light ratio so effectively that the model reproduces the observed luminosity dependence of clustering strength.
However, in the $\nu$GC model, the dust extinction has been adequately included in order to reproduce the observed color-magnitude relation and overall redshift distribution of each optical/NIR-selected galaxy sample \citep{nag05}.
Introducing more dust extinction in order to create the same strong dependence of mass on luminosity as observed would draw apparent contradictions with these observables, although the dust extinction in the $\nu$GC is treated by a simple model whose accuracy is still uncertain.
Real LBGs are expected to have a tighter correlation between halo mass and UV luminosity than expected from the semianalytic model prediction; however, it is not yet clear how to reproduce the tight correlation in the collisional starburst model, in which luminosity depends on the star formation history triggered by merging in a short period.
The scatter in the relation of halo mass and UV luminosity also significantly affects the shape of the luminosity function.
Model improvements are therefore required in order to reproduce both the observed correlation and luminosity functions simultaneously.




\subsection{Comparison with Previous work}

Our new and more reliable measurements of the ACF for LBGs have revealed that more luminous LBGs have steeper ACFs especially at small scales.
A hint of this signal was previously detected by \citet{ouc01} for LBGs at $z=4$.
On the contrary, \citet{all05} did not find such an ACF steepening for LBGs at $z=4$; however, their ACF could be measured only on scales larger than $10^{\prime \prime}$, so their results may be due to preferentially detecting the brightest central galaxy in a halo with their wide and shallow ($i<24.5$) data.
The steepening of the ACF for more luminous galaxies at $z=3$ was predicted by \citet[see their Fig.$12$]{kra04}, which suggests that this is due to a more pronounced one-halo component dominated by galaxy pairs within a single halo.
Our results of observation and the comparisons with a semianalytic model for luminous and massive LBGs at $z=4$ and $5$ are qualitatively in agreement with the implications of their work.
However, if such a steepening on small scales is a universal trend in high-$z$ galaxies, the clustering analysis made on a small contiguous FOV would detect only their one-halo components, leading to an incorrect measurement for the overall clustering strength at the epoch.
The actual prediction of \citet{kra04} is not the ACF steepening but the departure from a single power law of the spatial correlation function.
Based on deep imaging data, we have so far obtained only the ACF, which is the projection of the spatial correlation function, and therefore the steepening that we observe could be a result of dilution of their predicted trend.
Our present observational data have insufficient statistical resolution to allow the detection of the departures from a single power law that \citet{kra04} predicted for the correlation functions of high-$z$ galaxies.
Such a departure can be seen in a more statistically robust LBG sample \citep{ouc05}.
A future large spectroscopic survey for high-$z$ LBGs will derive the precise spatial correlation function, in which we could easily verify the noticeable transition scale from the two-halo term to the one-halo term.

\subsection{Dependance of Clustering on Age of Population}

Although we have shown that the observed ACF slope difference can be accounted for by LBG multiplicity in a single halo, here we consciously discuss an alternative possible interpretation to explain the observational result.
An amplitude enhancement of the bright LBG ACF has certainly been observed on small scales within the typical virial radius of a dark halo; however, this could be a projection effect owing to seeing an excess of halo-halo pairs on more extended spatial scales.
Such an effect can be expected, especially when there is large-scale structure traced by bright LBGs almost along the line of sight.
Although the CDM theory predicts that seeds of more massive dark halos collapse earlier with inherent stronger clustering, there could be an additional effect irrespective of dark halo mass to enhance the clustering difference between bright and faint LBGs.
If the luminous LBGs are an older population than faint LBGs, the observed ACF difference could be explained as a bias dependency on formation epoch.
\citet{yos01} found a clear tendency that early-forming galaxy populations are distributed tightly with the underlying mass density field, while the late-forming population tends to be less concentrated.
This is basically caused by stochastic biasing, so the formation of late-forming galaxies is suppressed in overdense and high-temperature regions where it is hard to cool gas effectively.
The tendency for a rich environment to favor the early formation of galaxies can be seen in the protocluster regions of $z\sim2$ LBGs \citep{ste05}.
Similar numerical results have been obtained by \citet{bla99} and \citet{som01}, which explain the clustering dependence on galaxy color seen in the local universe represented by the morphology-density relation.
The predicted correlation functions show obvious slope differences between old and young populations.
The same explanation could be applied to our finding of ACF slope differences in LBGs.
The essential assumption in this interpretation is that brighter LBGs should be older, which seems to be plausible from the observational implication that UV-luminous LBGs at $z=3$ have larger stellar masses \citep{pap02}.
However, \citet{sha01} found a contradictory tendency in that the UV luminosity is uncorrelated with the inferred stellar mass for the brighter LBG sample.
On the other hand, \citet{sha01} found a possible correlation between dust extinction and age for $z\sim3$ LBGs, although this depends strongly on the applied attenuation law.
If this is the case, our finding of no clustering difference between different $E(B-V)$ subsamples as seen in \S~5.3 could be an argument against the hypothesis described above that the age difference causes the ACF slope difference.
More studies are required to support this hypothesis.

\section{Conclusions}

We explored the clustering properties of LBGs at $z=4$ and $5$ with an angular two-point correlation function on the basis of the very deep and wide Subaru Deep Field data.
We constructed mock LBG catalogs based on a semi-analytic approach to the hierarchical clustering model combined with a high-resolution $N$-body simulation, carefully mimicking the observational selection effects.
The luminosity function for LBGs at $z=4$ and $5$ predicted by this mock catalog was found to be consistent with the observations.
Our main conclusions about the LBG clustering properties are as follows.

1. We confirmed the previous results that the clustering strength of LBGs depends on the UV luminosity range of the sample; that is, brighter LBGs are more strongly clustered.

2. We found apparent slope differences among UV luminosity subsamples of LBGs at both $z=4$ and $5$. 
More luminous LBGs have steeper correlation functions.
The bias parameter, defined as the ratio of the clustering strength of galaxies to that of dark matter, was found to be a scale-dependent function for bright LBGs, while it was almost scale-independent for faint LBGs.
Luminous LBGs have a high bias at smaller angular scales, which decreases as the scale increases.

3. The overall clustering strengths of LBGs at $z=4$ and $z=5$ are reproduced reasonably well by the hierarchical clustering model when the observational selection function is taken into account.
The observed clustering difference according to UV luminosity is not apparent in the model.
However, the halo-mass distinct samples were able to reproduce this difference.
More massive LBGs have steeper correlation functions.
The model predicts that LBG multiplicity in a massive dark halo makes this ACF slope difference, and halo occupation number plays an important role in the small-scale clustering of LBGs.

4. Our finding that there is a luminosity dependence of correlation function slope is probably an indication that massive dark halos must host multiple bright LBGs.
This is also supported by our semi-analytic model, which predicts that more massive dark halos have a higher occupation number of LBGs.

5. The hierarchical clustering model could reconcile the observed luminosity dependence of the ACF when there is a tight relation between UV luminosity and dark halo mass.

Our finding of a large difference in clustering strength on small scales will provide new insights into the substructure inside dark halos, on which we have unsolved questions: how many galaxies reside in a single halo, and how are they distributed and how do they evolve in a halo?
Interestingly, these questions are also raised by more precise measurements of clustering in the local universe, in which significant departures from a single power-law ACF on a small scale have been revealed \citep{zeh04, phl05}.
The nonlinear behavior of the bias parameter depending on scale is not surprising because the bias on small scales should depend on several complicated processes of interhalo dynamics, including cooling, star formation, supernova feedback, photoionization, merging, and tidal disruptions.
More precise measurements of galaxy clustering on small scales are required from upcoming observations of both the local and distant universe, and in addition, refinements of the semi-analytic approach with high resolution would help in better understanding the detailed galaxy formation and evolution processes in a dark halo.


\acknowledgments

We deeply appreciate the devoted technical and management support of the Subaru Telescope staff for this long-term project.
The observing time for this project was committed to all the Subaru Telescope builders.
We thank the referee for helpful comments that improved the manuscript.
The research is supported by the Japan Society for the Promotion of Science through Grant-in-Aid for Scientific Research 16740118.

\clearpage



\clearpage









\clearpage

\begin{deluxetable}{lrrrrr}
\tablecaption{ACF Parameters of SDF LBG Samples \label{tbl-acfparam}}
\tablewidth{0pt}
\tablehead{
\colhead{Sample} & \colhead{$N$\tablenotemark{a}} & \colhead{$f_c$(\%)\tablenotemark{b}} & \colhead{$\beta$\tablenotemark{c}} & \colhead{$A_w$\tablenotemark{c}} & \colhead{$r_0$\tablenotemark{d}}
}
\startdata
z4LBG total                 & 4543 & 2.6 & $-0.90_{-0.04}^{+0.06}$ &  $2.31_{-0.45}^{+0.44}$ & $4.69_{-0.67}^{+0.81}$\\
\tableline
z4LBG $i'\leq25.5$        &  916 & 0.38 & $-1.25_{-0.25}^{+0.19}$ & $10.44_{-4.36}^{+6.99}$ & $6.52_{-2.50}^{+4.09}$\\
z4LBG $25.5<i'\leq26.5$   & 1977 & 2.3 & $-0.98_{-0.12}^{+0.10}$ & $ 3.59_{-1.14}^{+1.58}$ & $5.38_{-1.44}^{1.97}$\\
z4LBG $26.5<i'\leq27.43$  & 1650 & 5.1 & $-0.80_{-0.16}^{+0.18}$ & $ 1.48_{-0.76}^{+1.07}$ & $4.09_{-1.66}^{+2.77}$\\
\tableline
z4LBG $(i'-z')>0.0$        & 1124 & - & $-1.08_{-0.10}^{+0.12}$ & $ 5.94_{-1.96}^{+2.00}$ & $6.14_{-1.58}^{+2.12}$\\
z4LBG $(i'-z')\leq0.0$     &  653 & - & $-1.04_{-0.66}^{+0.50}$ & $ 2.47_{-2.15}^{+7.08}$ & $4.19_{-3.05}^{+13.7}$ \\
\tableline
\tableline
z5LBG total                 &  831 & 10 & $-1.02_{-0.16}^{+0.14}$ & $ 7.16_{-2.33}^{+2.92}$ & $6.09_{-1.88}^{+2.66}$ \\
\tableline
z5LBG $z'\leq25.75$       &  325 & 9.0 & $-1.15_{-0.19}^{+0.15}$ & $18.68_{-5.92}^{+7.43}$ & $8.16_{-1.70}^{+3.76}$\\
z5LBG $25.75<z'\leq26.62$ &  506 & 11 & $-0.85_{-0.26}^{+0.27}$ & $ 3.03_{-1.77}^{+2.85}$ & $4.78_{-2.38}^{+5.24}$ \\
\enddata
\tablenotetext{a}{Number of sources in the sample.}
\tablenotetext{b}{Contamination rate rebinned in magnitudes from the original estimate by \citet{yos06} taking the average with number weighting.}
\tablenotetext{c}{The ACF power-law fitting parameters defined in Eq.(4) with contamination correction.}
\tablenotetext{d}{The correlation length inferred from ACF.}



\end{deluxetable}



\begin{deluxetable}{lrrrr}
\tablecaption{ACF Parameters of $\nu$GC LBG Samples \label{tbl-acfparamnugc}}
\tablewidth{0pt}
\tablehead{
\colhead{Sample} & \colhead{$N$\tablenotemark{a}} & \colhead{$f_c$(\%)\tablenotemark{a}} & \colhead{$\beta$} & \colhead{$A_w$} 
}
\startdata
z4LBG total  COL-selected  &12712 & 5.6 & $-0.74_{-0.06}^{+0.04}$ &  $0.83_{-0.11}^{+0.22}$ \\
z4LBG total  SF-selected   & 6653 & 0.0 & $-0.74_{-0.08}^{+0.08}$ &  $1.26_{-0.35}^{+0.33}$ \\
\tableline
z4LBG $i'\leq25.5$        &  4979 & 4.6 & $-0.80_{-0.04}^{+0.06}$ & $ 1.38_{-0.33}^{+0.28}$ \\
z4LBG $25.5<i'\leq26.5$   & 10735 & 6.4 & $-0.76_{-0.04}^{+0.02}$ & $ 0.76_{-0.10}^{+0.11}$ \\
z4LBG $26.5<i'\leq27.4$  & 11742 & 6.0 & $-0.82_{-0.02}^{+0.04}$ & $ 0.95_{-0.16}^{+0.10}$ \\
\tableline
z4LBG $12.0<$log$(M_h/M_\odot)$          & 2928 & 0.0 & $-1.39_{-0.01}^{+0.01}$ & $44.19_{-0.49}^{+0.49}$ \\
z4LBG $11.7<$log$(M_h/M_\odot)\leq12.0$  & 4556 & 0.0 & $-1.32_{-0.26}^{+0.06}$ & $13.32_{-1.14}^{+5.74}$ \\
z4LBG $11.5<$log$(M_h/M_\odot)\leq11.7$  & 5174 & 0.0 & $-0.92_{-0.04}^{+0.08}$ & $2.01_{-0.50}^{+0.39}$ \\
z4LBG $11.3<$log$(M_h/M_\odot)\leq11.5$  & 7201 & 0.0 & $-0.77_{-0.03}^{+0.11}$ & $0.69_{-0.26}^{+0.14}$ \\
z4LBG log$(M_h/M_\odot)\leq11.3$         & 5983 & 0.0 & $-0.64_{-0.11}^{+0.14}$ & $0.37_{-0.16}^{+0.21}$ \\
\tableline
z4LBG 1halo-1LBG $12.0<$log$(M_h/M_\odot)$   &  1796 & 0.0 & $-0.82_{-0.12}^{+0.12}$ & $2.88_{-1.22}^{+2.13}$ \\
z4LBG 1halo-1LBG $11.7<$log$(M_h/M_\odot)\leq12.0$ & 3783 & 0.0 & $-0.88_{-0.12}^{+0.12}$ & $2.00_{-0.85}^{+1.47}$ \\
z4LBG 1halo-1LBG $11.5<$log$(M_h/M_\odot)\leq11.7$ & 4931 & 0.0 & $-0.80_{-0.10}^{+0.12}$ & $0.96_{-0.45}^{+0.56}$ \\
z4LBG 1halo-1LBG $11.3<$log$(M_h/M_\odot)\leq11.5$ & 7135 & 0.0 & $-0.86_{-0.10}^{+0.10}$ & $0.87_{-0.32}^{+0.51}$ \\
z4LBG 1halo-1LBG log$(M_h/M_\odot)\leq11.3$  &  5960 & 0.0 & $-0.85_{-0.16}^{+0.19}$ & $0.67_{-0.37}^{+0.59}$ \\
\tableline
\tableline
z5LBG total COL-selected   &  2603 & 0.5 & $-0.81_{-0.07}^{+0.09}$ & $ 2.06_{-0.48}^{+0.45}$ \\
z5LBG total SF-selected    &  1690 & 0.0 & $-0.86_{-0.22}^{+0.18}$ & $ 2.56_{-1.25}^{+1.80}$ \\
\tableline
z5LBG $z'\leq25.75$       &  1611 & 0.12 & $-0.84_{-0.10}^{+0.12}$ & $ 3.17_{-1.20}^{+1.40}$ \\
z5LBG $25.75<z'\leq26.6$  &  3101 & 1.2 & $-0.74_{-0.06}^{+0.06}$ & $ 1.32_{-0.36}^{+0.42}$ \\
\tableline
z5LBG $12.1<$log$(M_h/M_\odot)$          &  440  & 0.0 & $-1.24_{-0.06}^{+0.08}$ & $49.70_{-8.01}^{+7.85}$ \\
z5LBG $11.8<$log$(M_h/M_\odot)\leq12.1$  &  1117 & 0.0 & $-1.04_{-0.06}^{+0.06}$ & $13.40_{-2.44}^{+1.73}$ \\
z5LBG $11.5<$log$(M_h/M_\odot)\leq11.8$  &  2547 & 0.0 & $-0.90_{-0.08}^{+0.08}$ & $2.40_{-0.74}^{+0.91}$ \\
\tableline
z5LBG 1halo-1LBG $12.1<$log$(M_h/M_\odot)$  &  385 & 0.0 & $-0.84_{-0.28}^{+0.30}$ & $7.59_{-5.77}^{+14.3}$ \\
z5LBG 1halo-1LBG $11.8<$log$(M_h/M_\odot)\leq12.1$  &  1029 & 0.0 & $-0.64_{-0.14}^{+0.18}$ & $2.09_{-1.18}^{+1.89}$ \\
z5LBG 1halo-1LBG $11.5<$log$(M_h/M_\odot)\leq11.8$  &  2486 & 0.0 & $-0.72_{-0.10}^{+0.18}$ & $1.33_{-0.72}^{+0.86}$ \\
\enddata
\tablenotetext{a}{$N$ and $f_c$ are average values for the random resampling process for the total and luminosity subsamples.}



\end{deluxetable}

\end{document}